\documentclass[letterpaper,12pt,margin=1in]{article}

\usepackage{graphicx}                
\usepackage{float}                   
\usepackage{hyperref}                
\usepackage[margin=1in]{geometry}  
\usepackage[short]{datetime}         
\usepackage[labelfont=bf]{caption}   
\usepackage[bottom]{footmisc}        
\usepackage{lineno}
\usepackage{ragged2e}
\usepackage[protrusion=true,expansion=true]{microtype} 
\usepackage[T1]{fontenc}                               
\usepackage{lmodern}

\usepackage[compact]{titlesec}
\titleformat*{\section}{\Large\bfseries}
\titlespacing*{\section}{0pt}{1ex}{1ex}
\titleformat*{\subsection}{\large\bfseries}
\titlespacing*{\subsection}{0pt}{1ex}{1ex}
\titleformat*{\subsubsection}{\large\bfseries}
\titlespacing*{\subsubsection}{0pt}{1ex}{1ex}

\usepackage{amsfonts,amssymb,amsmath, mathtools}
\usepackage{xcolor, color, soul}
\usepackage{graphicx}
\usepackage{multicol}
\usepackage{url}
\usepackage{faktor}
\usepackage{ulem}
\usepackage{pdfpages}
\usepackage{soul}
\usepackage{hyperref}
\usepackage{bm}
\usepackage{natbib}
\usepackage{subcaption}
\usepackage{booktabs}
\usepackage{setspace}
\usepackage{amsthm}
\usepackage{algorithm} 
\usepackage{algpseudocode}

\usepackage{caption}
\usepackage{array}
\usepackage{rotating}
\usepackage{xr}

\allowdisplaybreaks

\title{\normalsize{Model-Assisted Bayesian Estimators of Transparent Population Level Summary Measures for Ordinal Outcomes in Randomized Controlled Trials}}
\date{}

\begin{document}

\setstretch{1.2}





\center{\Large\textbf{Model-Assisted Bayesian Estimators of Transparent Population Level Summary Measures for Ordinal Outcomes in Randomized Controlled Trials} \\
\vspace{0.15in}
\normalsize{Lindsey E. Turner$^{1,*}$, Carolyn T. Bramante$^{2}$, and Thomas A. Murray$^{1}$ \\
1. Division of Biostatistics and Health Data Science, School of Public Health, University of Minnesota \\
2. Division of General Internal Medicine, University of Minnesota Medical School}}
\vspace{0.25in}
\begin{abstract}
In randomized controlled trials, ordinal outcomes typically improve statistical efficiency over binary outcomes. The treatment effect on an ordinal outcome is usually described by the odds ratio from a proportional odds model, but this summary measure lacks transparency with respect to its emphasis on the components of the ordinal outcome when proportional odds is violated. We propose various summary measures for ordinal outcomes that are fully transparent in this regard, including `weighted geometric mean' odds ratios and relative risks, and `weighted mean' risk differences. We also develop and evaluate efficient model-assisted Bayesian estimators for these population level summary measures based on non-proportional odds models that facilitate covariate adjustment with marginalization via the Bayesian bootstrap. We propose a weighting scheme that engenders appealing invariance properties, including to whether the ordinal outcome is ordered from best to worst versus worst to best. Using computer simulation, we show that comparative testing based on the proposed population level summary measures performs well relative to the conventional proportional odds approach. We also report an analysis of the COVID-OUT trial, which exhibits evidence of non-proportional odds.
\end{abstract}

%


\RaggedRight{\noindent\rule{2cm}{0.4pt} \\
*turne899@umn.edu}

\clearpage
\justifying

\setstretch{1.2}


\section{Introduction}

Randomized controlled trials (RCTs) are the gold standard for estimating causal effects of interventions on clinically relevant outcomes. The choice of primary outcome, and the approach for characterizing the treatment effect on this outcome, often reflects a compromise between clinical relevance, interpretability, and statistical efficiency. These choices should facilitate, not inhibit, drawing definitive and meaningful conclusions from the RCT \citep{popchak_framework_2019,iwashyna_choosing_2018}. Ordinal outcomes are useful because they can effectively distinguish between important aspects of the patient experience and often are more efficient than binary composites \citep{mchugh_simulation_2010,roozenbeek_added_2011}. In this way, they facilitate more quickly answering clinical questions about complex and evolving conditions \citep{selman_statistical_2024}. 

The COVID-OUT trial is one recent example where an ordinal outcome may be useful \citep{bramante_randomized_2022}. This trial investigated three re-purposed medications (metformin, ivermectin, and fluvoxamine) for outpatient treatment of SARS-CoV-2 infection to prevent progression to severe COVID-19 through 14 days (primary) and 28 days (secondary). The definition of severe COVID-19 was a 4-part composite including a reading of $\leq 93\%$ oxygen saturation on daily home oximetry, an emergency department (ED) visit, hospitalization, or death. A common approach to evaluating and characterizing the effect of a treatment on an outcome like this is with an odds ratio from a proportional odds (PO) model. However, when the treatment impacts components of the outcome in variable ways, or more precisely, when the PO assumption does not hold, this odds ratio lacks transparency regarding how it aggregates the break point specific treatment effects, making interpretation challenging. 

Previous work has relaxed this assumption using a partial proportional odds (PPO) and non-proportional odds models \citep{peterson_partial_1990, williams_richard_generalized_2006, fullerton_proportional_2012, schildcrout_model-assisted_2022}. 
Alternatively, a series of logistic regression models can provide estimates of break point specific treatment effects. 
While one could apply a joint hypothesis testing procedure for the set of treatment effects within a PPO modeling framework, such joint tests with potentially many degrees of freedom are inefficient and do not clearly specify whether rejecting the null hypothesis warrants an efficacy claim (e.g., the treatment may affect some categories positively and others negatively) \citep{bender_using_1998,peterson_partial_1990}. 
However, methods to aggregate the set of treatment effects are lacking, so many trials sacrifice efficiency in favor of transparency and interpretability by analyzing an ordinal outcome as a binary composite instead \citep{selman_statistical_2024}. Nevertheless, the treatment effect on a binary composite outcome still lacks nuance regarding treatment effects on each component of the outcome, which is particularly important when the components vary in incidence and clinical relevance. For example, while the COVID-OUT trial failed to reject the null hypothesis with respect to the binary composite primary outcome, this was driven by the lack of effect on the most frequently occurring and least reliable component, i.e. $\leq 93\%$ oxygen saturation on home oximetry tests. In pre-specified secondary analyses, there was nominally statistically significant evidence that metformin prevented the more reliable and clinically important healthcare utilization measures of severe COVID-19 \citep{bramante_randomized_2022}. 

In this paper, we aim to develop new transparent population level summary measures for characterizing treatment effects on ordinal outcomes with corresponding model-assisted Bayesian estimators. Our approach will facilitate alignment with the estimand framework, which suggests that treatment evaluations be based on a clearly defined population level summary measure \citep{kahan_estimands_2024}. 
\cite{neugebauer_nonparametric_2007} recently proposed a nonparametric extension used by \cite{diaz_enhanced_2016} to estimate the treatment effect on an ordinal outcome characterized as the average of the cumulative log odds ratios. Covariate adjusted estimators of this population level summary measure can provide further efficiency gains \citep{benkeser_improving_2021, williams_optimising_2022}. Although this approach is transparent, an average of the cumulative log odds ratio may be inefficient in scenarios where the outcome distribution is non-uniform. 
To address this limitation, we propose a general class of population level summary measures that reflect weighted geometric mean odds ratios or risk ratios, and weighted average risk differences. We consider various weighting schemes, and propose a particular weighting scheme that engenders various appealing invariance properties, including invariance to the outcome ordering (i.e., inference is not impacted by whether the outcome is ordered from best to worst versus from worst to best). We also develop efficient model-assisted Bayesian estimators for our proposed summary measures based on partial proportional odds models, which critically facilitates further potential efficiency gains through covariate adjustment and non-parametric marginalization with respect to the posterior covariate distribution via the Bayesian bootstrap  \citep{benkeser_improving_2021, rubin_bayesian_1981}. 

The rest of the paper is organized as follows. First, in Section \ref{sec:methods}, we define the proposed population level summary measures and detail our efficient model-assisted Bayesian estimators. In Section \ref{sec:sim}, using computer simulations, we evaluate the power of our proposed approach compared to other established methods, including the more conventional proportional odds model approach. Next, in Section \ref{sec:covid}, we apply our model-assisted, covariate-adjusted estimator of our proposed summary measures to re-evaluate the treatment effects on the primary outcome of the COVID-OUT trial. We conclude the paper with a brief discussion in Section \ref{sec:discussion}.

\section{Methods} \label{sec:methods}

\subsection{Population Level Summary Measures}

Let $Y$ denote an ordinal random variable taking values in $\mathcal{Y} = \{1,\ldots,K\}$ and $A$ denote the random treatment assignment in a two-arm randomized controlled trial with a control ($A=c$) arm and a treatment ($A=t$) arm. Without loss of generality, we assume $Y$ is encoded so that smaller values are better, i.e. $Y=1$ reflects the best health status and $Y=K$ reflects the worst health status. Let the probabilities of $Y$ be denoted as 
\begin{equation*}
    \pi_{k,a} = P(Y = k | A = a) \;\; \mathrm{and} \;\; \pi^{+}_{k,a} = \sum_{\ell=1}^k \pi_{\ell,a} = P(Y \leq k | A = a)
\end{equation*}
\noindent
for $k = 1,\ldots,K$ and $a \in \{c, t\}$. 
The treatment effect on an ordinal outcome $Y$ may be fully characterized by cumulative summary measures across break points $k=1,\ldots,K-1$, e.g., 
\begin{align*}
\text{OR}_k &= \{\pi^+_{k,t}/(1-\pi^+_{k,t})\} / \{\pi^+_{k,c}/(1-\pi^+_{k,c})\}, \\
\text{RD}_k &= \pi^+_{k,t} - \pi^+_{k,c}, \\
\text{RR}^{+}_k &= \pi^+_{k,t} / \pi^+_{k,c} \;\; \mathrm{or} \;\; \text{RR}^{-}_k = (1-\pi^+_{k,c}) / (1-\pi^+_{k,t}).
\end{align*}
\noindent
To ensure clear decision-making and improve efficiency, the fundamental challenge lies in aggregating these $K-1$ treatment effect summary measures into a univariate population level summary measure that is appropriate for characterizing the treatment effect such that sufficiently large positive values would warrant a superiority claim and sufficiently large negative values an inferiority claim. 
This aggregation is an inherently subjective exercise as there may be some $k \in \{1,\ldots,K-1\}$ such that the treatment is beneficial, i.e. $\text{OR}_k > 1, \text{RD}_k > 0, \text{RR}^+_k > 1, \; \mathrm{or} \; \text{RR}^-_k < 1$, and some $k^\prime \ne k$ where the treatment is not beneficial, i.e. $\text{OR}_{k^\prime} < 1, \text{RD}_{k^\prime} < 0, \text{RR}^+_{k^\prime} < 1, \; \mathrm{or} \; \text{RR}^-_{k^\prime} > 1$. 

Perhaps the most common aggregation approach is to use the odds ratio from a proportional odds (PO) model, which makes the PO assumption, i.e. $\text{OR}_1 = \ldots = \text{OR}_{K-1}$. In contrast, \cite{diaz_enhanced_2016} proposed summarizing the treatment effect as the average of the log ORs, which after exponentiation reflects the geometric mean OR, i.e. $\text{AOR} = \exp\lbrace\frac{1}{K-1} \sum_{\ell = 1}^{K-1} \log{\text{OR}_\ell}\rbrace$. Under proportional odds, these approaches reflect the same population level summary measure, i.e. the common OR. However, when the PO assumption fails, the OR from a PO model lacks transparency regarding how it aggregates these treatment effects as it is a maximum likelihood estimator that does not have an analytic solution to the score equations. The average log OR is transparent and logical when the incidence of the outcome distribution is uniform, but equal weights may be inefficient otherwise. Another option is the binary composite approach that focuses on the treatment effect at the first break point, e.g., $\text{OR}_1$, however this is myopic and often inefficient. 

To address the limitations of current approaches, we propose four classes of alternative population level summary measures for describing the effect of treatment relative to control on an ordinal outcome, which arise as follows:
\begin{align}
wOR &= \exp \left[ \sum_{\ell = 1}^{K-1} \left\{ w_\ell * \log\left( \text{OR}_{\ell} \right)\right\} / \sum_{\ell = 1}^{K-1} w_\ell \right], \\
wRD &= \sum_{\ell = 1}^{K-1} \left\{w_\ell * \left( \text{RD}_{\ell}\right)\right\} / \sum_{\ell = 1}^{K-1} w_\ell, \\
wRR^{+} &= \exp \left[ \sum_{\ell = 1}^{K-1} \left\{ w_\ell * \log\left( \text{RR}^{+}_{\ell}\right) \right\} /  \sum_{\ell = 1}^{K-1} w_\ell \right], \; \mathrm{and} \\
wRR^{-} &= \exp \left[ \sum_{\ell = 1}^{K-1} \left\{ w_\ell * \log\left( \text{RR}^{-}_{\ell}\right) \right\} /  \sum_{\ell = 1}^{K-1} w_\ell \right].
\end{align}
The $\{w_\ell\}_{\ell=1}^{K-1}$ in (1)-(4) reflect a general set of weights that we will discuss in detail below. 
That is, we propose weighted geometric mean odds ratios ($wOR$), weighted mean risk differences ($wRD$), and two versions of weighted geometric mean relative risks ($wRR^+$ and $wRR^-$). Because relative risk is not order invariant (i.e. $\log\{\pi^+_{k,t}/\pi^+_{k,c}\} \ne -\log\{(1-\pi^+_{k,t})/(1-\pi^+_{k,c})\}$), we propose two weighted RR measures depending on if the RR reflects a risk ratio of positive outcomes ($RR^{+}$) versus negative outcomes ($RR^{-}$).
We propose to use geometric means for $wOR$ in (1) and $wRR$ in (3) and (4) as the cumulative OR and cumulative RR are asymmetric on the additive scale.


Regarding the weights, we propose to define
\begin{equation}\label{eq:weights}
w_k = (\pi_{k,c} + \pi_{k+1,c})\pi_{k,c}^{+}(1 - \pi_{k,c}^{+}), 
\end{equation}
for $k=1, \ldots, K-1$.
These weights have (or are expected to have) various appealing properties.
\cite{dg_clayton_odds_1974} derived weights that minimize variance under the null hypothesis for a summary log odds ratio of an ordinal outcome. These optimal weights arose as $w_k = (\pi_k + \pi_{k+1})\pi_k^{+}(1 - \pi_k^{+})$, where $\pi_k = (\pi_{k,t} + \pi_{k,c})/2$ and $\pi^+_k = (\pi^+_{k,t} + \pi^+_{k,c})/2$, for $k=1,\ldots,K-1$.
Because our proposed weights in (5) share the same multiplicative structure as these optimal weights, we expect our proposed weights to be more efficient than alternatives choices like uniform weights, i.e. $w_1 = \ldots = w_{K-1}$. 
In contrast to the optimal weights, our proposed weights are defined by the outcome distribution under the control rather than the average of the outcome distribution under the control and the treatment. 
As such, our proposed weights are invariant to the treatment effect.
This is an appealing property because, for example, unlike the optimal weights, our proposed weights will not reduce the weight assigned to $(K-1)$-th effect (e.g., $\text{OR}_{K-1}$ in (1)) when the treatment being evaluated reduces the incidence of the $K$-th outcome. 
Moreover, we chose to focus on the control distribution because in a randomized controlled trial of an experimental treatment, the trial population is currently receiving the control. Thus, weighting on the distribution of the control naturally reflects the general population if they were not in the RCT. 


There are two additional appealing properties shared by our proposed weighting scheme in (5) and the optimal weighting scheme derived in \cite{dg_clayton_odds_1974}. First, in a PO model, the OR is order invariant in the sense that $\text{OR}_k = 1/\text{OR}_{k+1}$ when the outcome is reverse encoded (i.e. $Y=1$ reflects the worst health status and $Y=K$ reflects the best health status). The OR from a PO model also shares this invariance property.  
Although using $w_k = \pi_{k}$ may seem reasonable, this would reflect the arbitrary choice of ordering from best to worst rather than worst to best. When the outcome is ordered from best to worst, the weighted average would be $\sum_{\ell = 1}^{K-1} \pi_{\ell, a} \log(\text{OR}_\ell) / \sum_{\ell = 1}^{K-1} \pi_{\ell, a}$. In contrast, when the outcome is ordered from worst to best, the weighted average would be $\sum_{\ell = 1}^{K-1} \pi_{\ell+1, a} \log(1/\text{OR}_{\ell + 1}) / \sum_{\ell = 1}^{K-1} \pi_{\ell + 1,a}$. These two statements are only equivalent when $\pi_{k,a} = \pi_{k+1,a}$, for all $k \in {1, \ldots, K}$, i.e. when the outcome distribution in arm $a$ is uniform. These same arguments hold for the proposed weighted risk difference measure defined in (2), but not for the weighted relative risk measures defined in (3) and (4) due to the lack of order invariance for univariate relative risk measures. 

The proposed weighting scheme in (5) also makes the corresponding population level summary measures invariant to the addition of a level of the outcome with probability zero. To illustrate this, consider a simple example of a three level ordinal outcome where the distribution of the outcome in the control arm is (0.33, 0.33, 0.33) and the distribution in the treatment arm is (0.50, 0.25, 0.25). Then $\text{OR}_1 = 2$ and $\text{OR}_2 = 1.5$. Since the control arm is evenly distributed, both AOR and $wOR$ are 1.73. Suppose we arbitrarily add a component with zero probability between previous levels 2 and 3 such that the control arm is distributed as (0.33, 0.33, 0.00, 0.33) and the treatment arm is distributed as (0.50, 0.25, 0.00, 0.25). Now, $\text{OR}_1 = 2$, $\text{OR}_2 = 1.5$, and $\text{OR}_3 = 1.5$. In this scenario, the AOR is 1.65, whereas $wOR$ using the proposed weights in (5) is still 1.73. When a new level is added arbitrarily, using equal weights changes the population level summary measure while our proposed approach remains unchanged. This property holds regardless of where a new level with probability is added, including at the lowest and highest levels. Additionally, the weights put a higher priority on the OR, RD, and RR for more frequent events and lower priority on any events that are less common. Although an unweighted average is order invariant, it is sensitive to how the ordinal outcome is defined and gives equal weight to components that rarely occur and components that occur frequently. Our proposed method takes into account that the ordinal outcome may not be evenly distributed.  

\subsection{Model-Assisted Bayesian Estimators}
To facilitate estimation of the population level summary measures defined in (1) - (4) with the proposed weights defined in (5), we propose to use a Bayesian marginal partial proportional odds (PPO) model \citep{peterson_partial_1990}. The PPO model allows a different treatment effect at each break point $k=1,\ldots,K-1$. In contrast, the PO model requires the treatment effect to be the same at each break point such that the odds ratio for the treatment effect is constant. In this way, the PPO model allows for non-proportional odds and more flexibility than the PO model. 

Without baseline covariates, the PPO model assumes
\begin{equation}
P(Y \leq k | A, \alpha_k, \beta, \tau_k) = \text{expit}\lbrack\alpha_k + \beta \lbrace I(A=t)-0.5 \rbrace + \tau_{k-1} \lbrace I(A=t)-0.5 \rbrace \lbrace I(k \geq 2) \rbrace \rbrack, 
\end{equation}
for $k = 1, \ldots, K-1$, where $\text{expit}(x) = e^x/(1+ e^x)$ and $I(Z)$ is a binary indicator for event $Z$.
We assume weakly informative priors
\begin{align*}    
\text{expit}(\bm{\alpha}) & \sim \text{Dirichlet}\left(\frac{1}{0.8 + 0.35 \max(K,3)}\right), \\
\beta & \sim \text{Normal}(\mu = 0, \sigma = 100),  \;\; \mathrm{and} \\
\tau_k & \stackrel{ind.}{\sim} \text{Normal}(\mu = 0, \sigma = 100), \; k=1,\ldots,K-2,  
\end{align*}
where $\bm{\alpha} = (\alpha_1,\ldots,\alpha_{K-1})$.
The concentration parameter of the Dirichlet distribution has been chosen so the posterior mean of the intercepts is very similar to the MLE \citep{rmsb}.

Because the proposed population level summary measures and weighting schemes are functions of the outcome distribution probabilities $\pi_{k,c}$ and $\pi_{k,t}$, we may simply transform posterior samples for the model parameters (i.e., $\bm{\alpha}$, $\bm{\beta}$) to obtain posterior samples for these population level summary measures. For example, under the assumed PPO model,
\begin{align*}
    \pi_{1,c} &= \text{expit}(\alpha_1 - 0.5\beta),  \\
\pi_{2,c} &= \text{expit}(\alpha_2 - 0.5\beta - 0.5\tau_1) - \text{expit}(\alpha_1 - 0.5\beta),\\
\pi_{k,c} &= \text{expit}(\alpha_k - 0.5\beta - 0.5\tau_{k-1}) - \text{expit}(\alpha_{k-1} - 0.5\beta - 0.5\tau_{k-2}), \; \mathrm{for} \; k = 2, \ldots,K-1,  \\
\pi_{K,c} &= 1-\text{expit}(\alpha_{K-1} - 0.5\beta_{K-1} - 0.5\tau_{K-2})
\end{align*}

\subsection{Covariate Adjustment}

Thus far, we have proposed to estimate our marginal population level summary measures using a marginal PPO model without baseline covariates. We now extend our estimator to facilitate covariate adjustment, which generally improves efficiency \citep{benkeser_improving_2021, williams_optimising_2022}. Additionally, in \cite{dg_clayton_odds_1974}, they propose an analysis for stratification, but do not allow for covariate adjustment. 
To estimate the proposed marginal summary measures with a non-collapsible adjusted model, we will use Bayesian g-computation coupled with the Bayesian bootstrap to account for uncertainty in the population covariate distribution \citep{willard_covariate_2024, rubin_bayesian_1981}. In our adjusted model, the marginal $\pi_{k,a}$ will be derived as  
\[\pi_{k,a} = P(Y = k| A= a) = \int P(Y = k| A = a, \bm{X}) dF_{\bm{X}}. \]

To estimate these $\pi_{k,a}$, we fit a Bayesian PPO model adjusting for covariates $\bm{X} = (X_1,\ldots,X_p)$, and then estimate $F_{\bm{X}}$ using the Bayesian bootstrap.
Similar to the unadjusted model, we use a Bayesian PPO model that includes $\sum_{\ell = 1}^p \gamma_\ell X_\ell$ component in the linear predictor of equation (6),
where we use the same prior distributions for $\bm{\alpha}$, $\beta$, and $\bm{\tau}$ as before and assume $\gamma_{\ell} \stackrel{ind.}{\sim} \text{Normal}(\mu = 0, \sigma = 100)$, $\ell = 1, \ldots, p$. 
After fitting the model, we use the Bayesian bootstrap by drawing weights from a Dirichlet distribution for each observation and taking the weighted average to estimate $\pi_{k,c}$ and $\pi_{k,t}$ from the Bayesian PPO model \citep{rubin_bayesian_1981}. 
Ultimately, we use $\pi_{k,c}$ and $\pi_{k,t}$ to estimate $wOR$, $wRD$, $wRR^{+}$, and $wRR^{-}$.




\section{Simulation Study} \label{sec:sim}

\subsection{Data Generative Scenarios}

We conducted a simulation study to investigate the type I error and power of comparative testing procedures based on the proposed metrics compared to established methods, including a PO model approach and the net benefit \citep{buyse_net_2021}. We estimate the weighted and unweighted average metrics with a Bayesian PPO model and the common odds ratio with a Bayesian PO model. We also evaluate the break point specific odds ratios with logistic regression models that treat the outcome as a binary composite at the corresponding break point. To compare Bayesian and frequentist methods, we defined ``statistical significance" as the 95\% confidence interval (binary composite outcomes and net benefit) or credible interval (all other metrics) excluding the null value corresponding to no difference between the randomization groups. For each scenario described below, we calculated the number of times each method was statistically significant in 2,000 replicate datasets with 1,000 participants. 

We considered an ordinal outcome with $K=5$ categories (e.g., nothing (1), hypoxemia (2), ED visit (3), hospitalization (4), death (5)). 
We generated datasets under two trial populations with differing control arm distributions and for each population we 
considered nine treatment arm distributions exhibiting various proportional and non-proportional odds behavior. 
We did not consider covariate adjustment in our computer simulation study.

\begin{figure}
 \centerline{\includegraphics[width=6in]{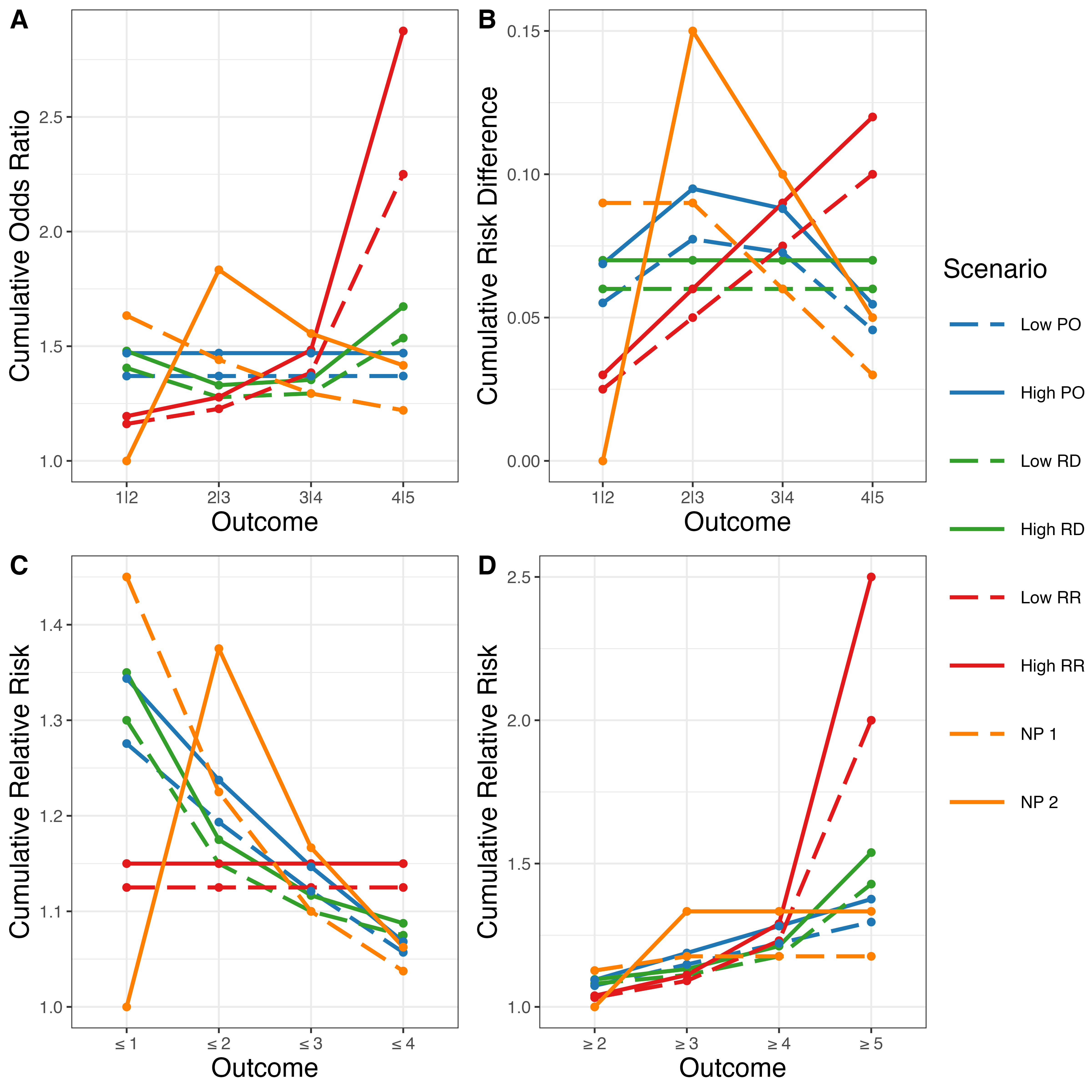}}
\caption{Simulation scenarios for the evenly distributed ordinal outcome (Setting 1). Panels A, B, C, and D show the cumulative OR, cumulative RD, cumulative $\text{RR}^{+}$, and cumulative $\text{RR}^{-}$, respectively, for each of the outcome levels.}
 \label{f:simscen1}
\end{figure}

In the first population, we consider the control has equal probability (20\%) of each of the five events in the outcome. 
For the treatment arm, we consider nine different treatment effects depicted in Figure~\ref{f:simscen1}, which shows the break point specific effects as cumulative OR, RD, $\text{RR}^{+}$, and $\text{RR}^{-}$. The corresponding distributions of the control and treatment arms are displayed in Figure S1. In scenario 1, there is a null treatment effect, all events have an equal probability. Scenarios 2 (low PO) and 3 (high PO) have a low (1.36) and high (1.48) constant OR. These two scenarios meet the PO assumption and were created to give approximately 80\% and 92\% power for 1,000 participants with a PO model. In scenarios 4 (low RD) and 5 (high RD), we will consider a constant cumulative RD. Scenario 4 has a lower RD of 0.06, while scenario 5 has a slightly higher RD of 0.07. Scenarios 6 (low RR) and 7 (high RR) will have a low (1.125) and high (1.15) constant cumulative $\text{RR}^{+}$. The final two scenarios are nonparametric scenarios. Scenario 8 (NP 1) has a 15\% reduction in the three worst outcomes (e.g., ED visit, hospitalization and death), while all of the benefit (45\% increase) goes to the best outcome (nothing bad). Scenario 9 (NP 2) has a 25\% reduction in the three worst outcomes (e.g., ED visit, hospitalization, and death), while the second best outcome (e.g., hypoxemia) will receive the benefit (75\% increase). In this scenario, there is no difference in the proportion of patients that get the best outcome. 

\begin{figure}
 \centerline{\includegraphics[width=6in]{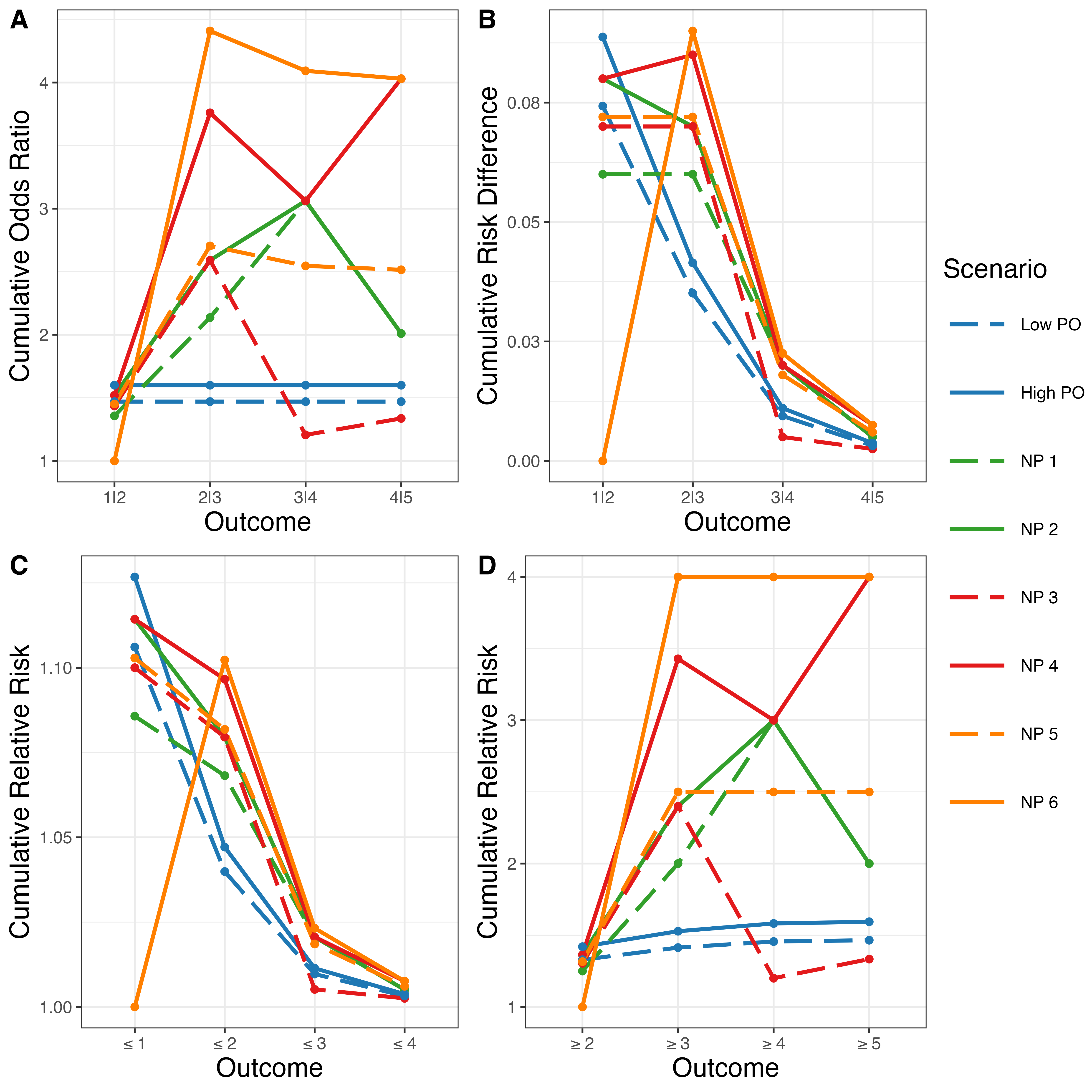}}
\caption{Simulation scenarios for the ordinal outcome distributed following the COVID-OUT outcome distributions. Panels A, B, C, and D show the cumulative OR, cumulative RD, cumulative $\text{RR}^{+}$, and cumulative $\text{RR}^{-}$, respectively, for each of the outcome levels.}
 \label{f:covidscen}
\end{figure}

We also considered a population where the control arm had a similar distribution as the COVID-OUT trial. The scenarios are depicted in Figure~\ref{f:covidscen}, and the control and treatment arm distributions are available in Figure S1. In this setting, the control arm is distributed with a large probability of the lowest level (70\% for nothing bad) and decreasing probabilities of the other levels (18\% for hypoxemia, 9\% for ED visit, 2\% for hospitalization, and 1\% for death). The two most extreme outcomes are very rare. Again, for the treatment arm, we will consider nine different treatment effects. Scenario 10 has a null treatment effect. Scenarios 11 (low PO) and 12 (high PO) follow PO with a low (1.47) and high (1.60) constant OR. These scenarios were created to give approximately 80\% and 92\% power, respectively, for 1,000 participants with a PO model. The next six scenarios (NP 1-6) represent various scenarios where the treatment effect varies across the outcomes. Due to the rareness of the most severe outcomes, in all these scenarios, the RD and $\text{RR}^{+}$ are decreasing as the severity of the outcome increases. However, as seen in panel A of figure~\ref{f:covidscen}, the ORs fluctuate as the outcome becomes more severe. Scenario 17 (NP 5) has a 50\% reduction in the three worst outcomes (ED visit, hospitalization and death), while the benefit of 10\% is added to the first outcome (nothing bad). Finally, scenario 18 (NP 6) has a 75\% reduction in the three worst outcomes (ED visit, hospitalization, and death), while the second best outcome (hypoxemia) has a 50\% increase. 

\subsection{Results}
Table~\ref{t:evenpower} displays the power for each of the different scenarios in the first population where the control arm is equally distributed across all levels of the outcome. Since there were 2,000 replicates for each simulation, we would expect the Monte Carlo error to be less than 2\% for the power estimates and about 0.5\% for the type I error rate estimates. 
In the null scenario, the type I error rate is relatively constant around 5\% for all measures. 
When the PO assumption holds, all ordinal composite measures have very similar power. 
In the scenarios where the PO assumption is violated, the proposed $wOR$ tends to have similar or higher power, specifically in the low RR and high RR scenarios, the $wOR$ has higher power than the PO model.
Despite the OR from the PO model not being defined, we estimated the true value using numerical techniques reported in Table S1. It is evident that $wOR$ is larger than the OR from the PO model, which is likely leading to some power gains. The $wOR$ with the Clayton weights are approximately equal to the OR from the PO model. Additionally, Table S2 shows further analysis from these simulation studies. Again, when the PO assumption is met, these proposed summary measures all have similar power. In the scenarios when the PO assumption is violated, the $wRD$ tends to have slightly lower power than the PO model besides in scenario NP 2. The $wRR^{-}$ tends to have higher power than the $wRR^{+}$ except in scenario NP 1. Finally, for the $wOR$, we looked at using each component of the weights separately. The $wOR$ sum defines the weights as $w_k = (\pi_{k,c} + \pi_{k+1,c})$, whereas the $wOR$ cumulative defines the weights as $w_k = \pi_{k,c}^{+}(1 - \pi_{k,c}^{+})$. Both of these measures tend to have similar power to the $wOR$ measure. 

We also dichotomized the outcome and fit a logistic regression model for the binary outcome. Overall, these binary outcomes tended to have much lower power. There was only one scenario (NP 1) where dichotomizing the outcome at the first break point had a higher power than analyses using the ordinal outcome. In this scenario, the treatment had an equal reduction in the three worst outcomes and all of the benefit was seen in the nothing bad outcome. As seen in Table ~\ref{t:evenpower}, there are certain cases where a dichotomization may provide more power than analyses using the ordinal outcome. This is when the combined treatment effect is optimized over the outcomes. For example, in the constant RR scenarios where the worst outcome has a decreased probability, but all other outcomes have a slightly higher probability, the binary logistic regression of the worst outcome has the highest power. However, there is no way of predicting this would be the optimal dichotomization.

\begin{table}
\centering
 \def\~{\hphantom{0}}

\resizebox{\columnwidth}{!}{%
\begin{tabular}{>
{\centering\arraybackslash}p{4em}cccccccc}
\toprule
& \multicolumn{4}{c}{Ordinal Composite} 
& \multicolumn{4}{c}{Binary Composite} \\
\cmidrule(l){2-5}\cmidrule(l){6-9}
Scenario & PO & NB & AOR & wOR & Nothing & Hypoxemia & ED & Hospitalized\\
\midrule
Null & 0.06 & 0.06 & 0.06 & 0.06 & 0.04 & 0.05 & 0.05 & 0.04\\
Low PO & 0.80 & 0.79 & 0.79 & 0.80 & 0.55 & 0.69 & 0.64 & 0.47\\
High PO & 0.92 & 0.92 & 0.92 & 0.93 & 0.74 & 0.85 & 0.81 & 0.61\\
Low RD & 0.75 & 0.75 & 0.80 & 0.76 & 0.62 & 0.47 & 0.49 & 0.71\\
High RD & 0.88 & 0.88 & 0.91 & 0.89 & 0.75 & 0.61 & 0.63 & 0.86\\
Low RR & 0.79 & 0.79 & 0.91 & 0.82 & 0.16 & 0.36 & 0.69 & 1.00\\
High RR & 0.93 & 0.92 & 0.98 & 0.95 & 0.21 & 0.49 & 0.85 & 1.00\\
NP 1 & 0.86 & 0.86 & 0.83 & 0.86 & 0.91 & 0.81 & 0.49 & 0.24\\
NP 2 & 0.92 & 0.92 & 0.87 & 0.91 & 0.04 & 1.00 & 0.91 & 0.56\\
\bottomrule
\end{tabular}}
  \caption{Power for Setting 1 - Evenly Distributed Outcome. The first four columns show the power of an ordinal outcome using a proportional odds model (PO), net benefit (NB), AOR, and $wOR$. The last four columns show the power of a binary composite outcome with logistic regression for each of the levels of the ordinal outcome.}
\label{t:evenpower}
\end{table}

In the nine scenarios where the population distribution mimics the COVID-OUT trial, there are very similar results in terms of power (Table~\ref{t:covidpower}). 
Again, the type I error rate is relatively constant around 5\%.
Comparing the AOR to the $wOR$, in most scenarios, the $wOR$ has a substantially higher power. When the PO assumption holds, the $wOR$ has similar power to both the PO model and the net benefit. When the PO assumption is not true, the $wOR$ has higher power than both the PO model and net benefit in all six scenarios. 
Additionally, Tables S3 and S4 show further results from these simulation studies. When the PO assumption holds, all three measures have similar power to the PO model. When the PO assumption does not hold, the $wRD$ and $wRR^{+}$ tend to have lower power than the PO model. However, in these scenarios, $wRR^{-}$ tends to have higher power than the PO model. The $wOR$ sum and $wOR$ cumulative also tend to have higher power than the PO model and net benefit when the PO assumption does not hold. 
There is a very similar trend as seen in the other setting in the dichotomized logistic regression models. The dichotomized logistic regression models tend to only have sufficient power when the outcome is dichotomized at the point that maximizes the treatment effect. When the PO assumption holds, all of the binary outcomes have a lower power than the ordinal outcomes. 


\begin{table}
\centering
 \def\~{\hphantom{0}}

\resizebox{\columnwidth}{!}{%
\begin{tabular}{>
{\centering\arraybackslash}p{4em}cccccccccccc}
\toprule
& \multicolumn{4}{c}{Ordinal Composite} 
& \multicolumn{4}{c}{Binary Composite} \\
\cmidrule(l){2-5}\cmidrule(l){6-9}
Scenario & PO & NB & AOR & wOR & Nothing & Hypoxemia & ED & Hospitalized\\
\midrule
Null & 0.04 & 0.04 & 0.06 & 0.04 & 0.05 & 0.05 & 0.05 & 0.05\\
Low PO & 0.78 & 0.77 & 0.30 & 0.79 & 0.76 & 0.46 & 0.17 & 0.07\\
High PO & 0.91 & 0.91 & 0.40 & 0.91 & 0.90 & 0.60 & 0.21 & 0.07\\
NP 1 & 0.71 & 0.71 & 0.60 & 0.75 & 0.57 & 0.93 & 0.65 & 0.11\\
NP 2 & 0.91 & 0.91 & 0.68 & 0.93 & 0.83 & 0.99 & 0.65 & 0.11\\
NP 3 & 0.82 & 0.82 & 0.38 & 0.85 & 0.71 & 0.99 & 0.08 & 0.05\\
NP 4 & 0.93 & 0.93 & 0.88 & 0.96 & 0.83 & 1.00 & 0.65 & 0.20\\
NP 5 & 0.86 & 0.86 & 0.73 & 0.89 & 0.73 & 0.99 & 0.53 & 0.14\\
NP 6 & 0.15 & 0.15 & 0.85 & 0.31 & 0.05 & 1.00 & 0.77 & 0.20\\
\bottomrule
\end{tabular}}
  \caption{Power for Setting 2 - Mimicking COVID-OUT. The first four columns show the power of an ordinal outcome using a proportional odds model (PO), net benefit (NB), AOR, and $wOR$. The last four columns show the power of a binary composite outcome with logistic regression for each of the levels of the ordinal outcome.}
\label{t:covidpower}
\end{table}

\section{COVID-OUT Analysis} \label{sec:covid}

We applied the proposed approach to the COVID-OUT trial. This trial randomized a total of 1,431 non-hospitalized persons with COVID-19 to either metformin or placebo, of which 1,323 initiated blinded study drug and were included in the modified-intent to treat analysis, where 663 initiated metformin and 660 initiated placebo. While the COVID-OUT trial also evaluated ivermectin and fluvoxamine as treatments using a 2x3 factorial design, we focus on the metformin evaluation here. We analyze the primary disease progression outcome as an ordinal outcome with four different levels: nothing bad (1), hypoxemia (2), ED visit (3), and hospitalization or death (4). We chose to combine hospitalization and death due to the very low incidence of death which caused convergence problems for the Bayesian PPO model. The primary outcome was measured through day 14, and secondarily through day 28. In the actual trial, this outcome was analyzed as a binary composite (i.e., hypoxemia, ED visit, hospitalization, or death) using logistic regression with adjustments for the other study drugs (ivermectin and fluvoxamine) and baseline vaccination status \citep{bramante_randomized_2022}. In this paper, we also adjust for other study drugs and vaccination status. 

\begin{figure}
 \centerline{\includegraphics[width=6in]{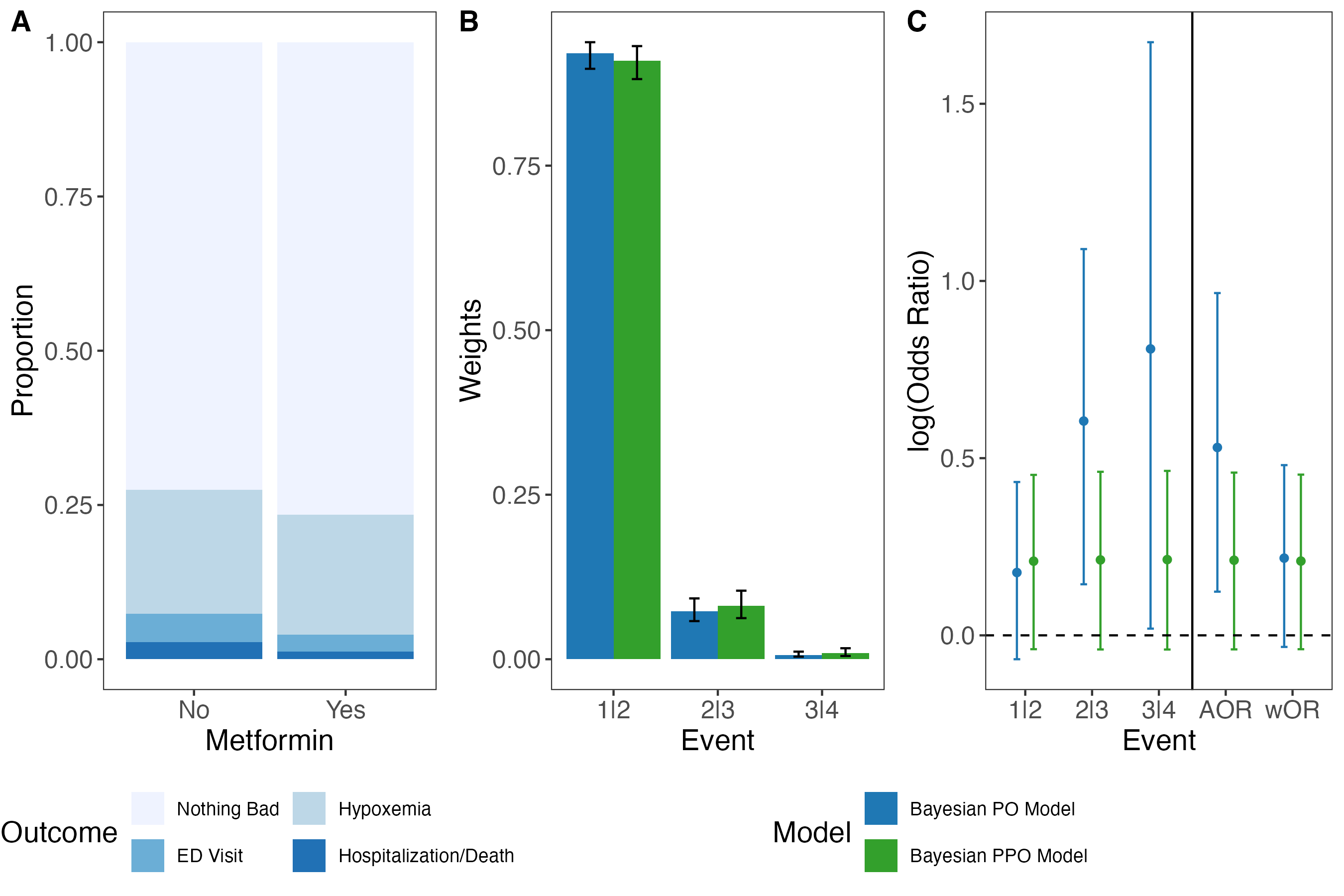}}
\caption{Adjusted analyses for COVID-OUT dataset at day 14 for metformin relative to placebo. Panel A shows the distribution of the outcome by treatment arm. Panel B shows the distribution of the weights. Panel C shows the log(OR). First showing the individual cumulative ORs when dichotomizing the outcome and then the AOR and $wOR$ for the ordinal outcome for both a Bayesian partial proportional odds (PPO) model (blue) and a Bayesian proportional odds (PO) model (green) using the Bayesian bootstrap for marginalization. The events are defined as nothing bad (1), hypoxemia (2), ED visit (3), and hospitalization/death (4).}
 \label{f:adj14}
\end{figure}

Figure~\ref{f:adj14} depicts the individual cumulative summary measures as well as the unweighted and weighted model-assisted weighted geometric mean odds ratio ($wOR$) estimates based on Bayesian PO and PPO models. Results for the weighted RD and RR measures are detailed in the Supporting Information. Panel A shows the distribution of the outcomes at day 14. The majority of the participants had nothing bad or hypoxemia, and few participants had an ED visit or hospitalization/death. 
Panel B shows the weights for each model-assisted estimate. The treatment effect on nothing bad (1) and hypoxemia (2) has the largest weight as these components exhibit the highest frequency. The other weights are both relatively small. 
Panel C shows the cumulative ORs for each level of the ordinal outcome, AOR, and $wOR$. In this panel, the PPO model demonstrates that the PO assumption may be violated as the cumulative OR is increasing as the outcome becomes more severe. Based on LOO CV metrics, the PO model and the PPO model fit the data similarly well. This provides moderate evidence that the PO assumption may not be met as if this assumption was met, we would expect the PO model to have a better fit.  
Notably, the credible interval for the estimate for the OR of ED visit or better is the largest and least precise.  
The AOR=1.70 (95\% CI: 1.13 to 2.63), whereas the proposed $wOR$=1.24 (95\% CI: 0.97 to 1.62).
The AOR estimate is larger than the $wOR$ estimate due to the AOR weights placing greater emphasis on the rarer outcomes which happen to also exhibit the larger treatment effect.
This differential emphasis of the rare outcomes also causes the AOR to have a 95\% CI that is 2.31 times wider than the proposed $wOR$.
The PO model assumes the conditional ORs across each break point is the same. 
However, due to non-collapsibility and the use of Bayesian g-computation to marginalize over the covariates, the marginal ORs from the PO model are not identical.
To provide a comparable measures for the PO model, we applied the proposed weights to obtain the PO model $wOR$=1.23 (95\% CI: 0.96 to 1.57).
The model-assisted wOR estimates based on the PPO and PO models are very similar.

\begin{figure}
 \centerline{\includegraphics[width=6in]{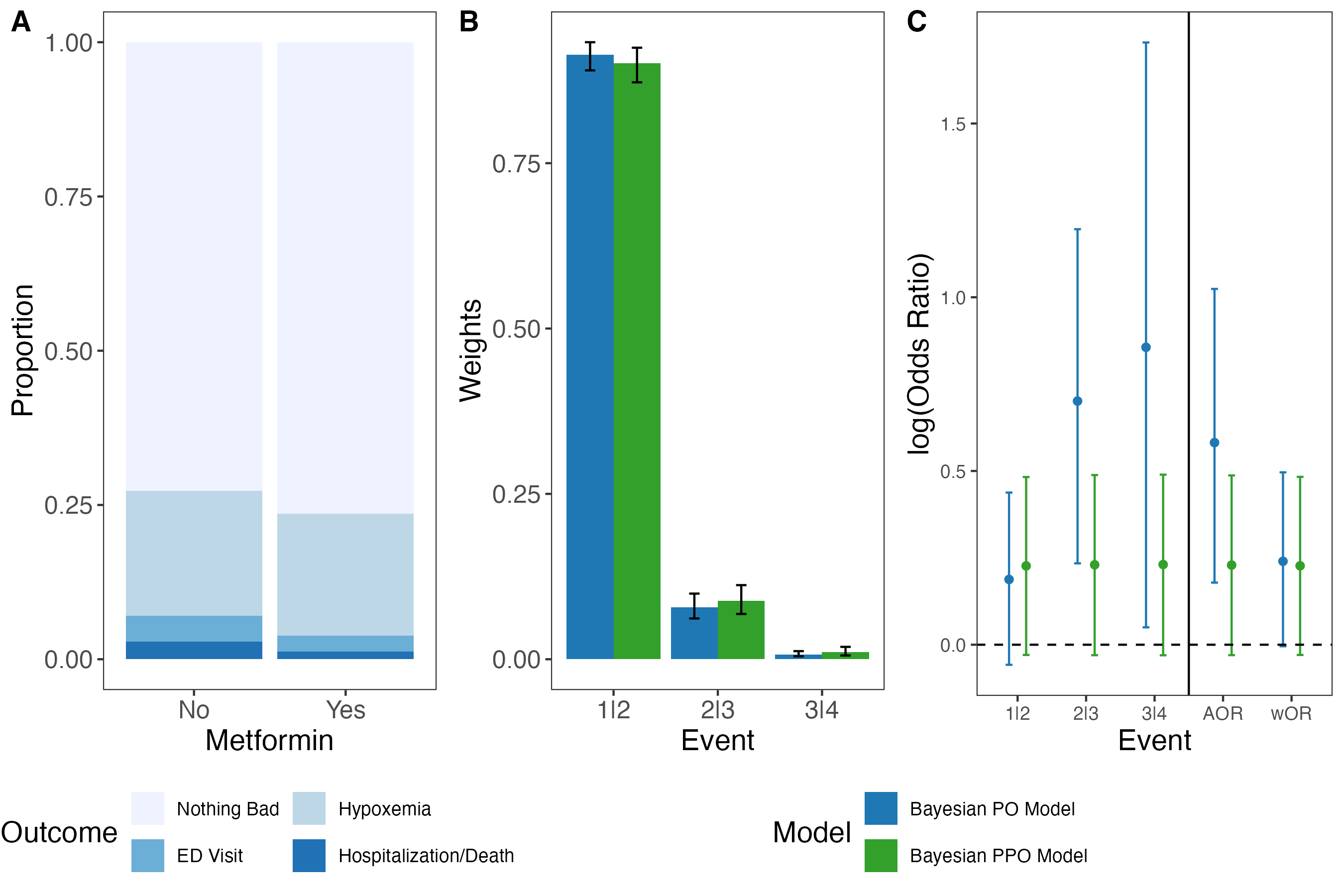}}
\caption{Adjusted analyses for COVID-OUT dataset at day 28 for metformin relative to placebo. Panel A shows the distribution of the outcome by treatment arm. Panel B shows the distribution of the weights. Panel C shows the log(OR). First showing the individual cumulative ORs when dichotomizing the outcome and then the AOR and $wOR$ for the ordinal outcome for both a Bayesian partial proportional odds (PPO) model (blue) and a Bayesian proportional odds (PO) model (green) using the Bayesian bootstrap for marginalization. The events are defined as nothing bad (1), hypoxemia (2), ED visit (3), and hospitalization/death (4).}
 \label{f:adj28}
\end{figure}

At day 28, as seen in Figure~\ref{f:adj28}, the trends in the treatment effect of metformin are similar. Again, there are very few participants that have the most extreme outcomes and the weight dichotomizing the ordinal outcome at nothing bad (1) and hypoxemia (2) is by far the largest. The OR is showing very similar trends of a slight benefit for metformin. 
All measures are showing a slight benefit for metformin, though only the equally weighted summary measures (e.g., AOR) would merit an efficacy claim (i.e., 95\% CI excluding the null value). 
The $wOR$ from the PPO model is 1.27 (95\% CI: 0.996 to 1.64) and the PO model-based $wOR$ is very similar at 1.26 (95\% CI: 0.97 to 1.62), but again while there is not strong evidence from LOO CV that the PPO model fits the data better than the PO model, there are still concerns about the validity of the PO assumption being met.

Briefly, for the PPO model, the effect is similar for each of the three individual cumulative RDs at day 14 (Figure S2). All RDs are showing a slight benefit towards metformin compared to the control. The credible interval for the ARD (0.03, 95\% CI: 0.004 to 0.05) excludes the null value, while the credible interval for the $wRD$ (0.03, 95\% CI: -0.01 to 0.07) does not. Both of these indicate about a 3\% increase in absolute risk of a bad outcome in the control group compared to the metformin group. For the PO model, the cumulative risk difference decreases as the outcome is more extreme. Both the ARD (0.02, 95\% CI: -0.003 to 0.04) and $wRD$ (0.03, 95\% CI: -0.01 to 0.07) credible intervals include the null value.
In panel B, the cumulative $\text{RR}^{+}$s for the individual outcomes, $\text{ARR}^{+}$, and $wRR^{+}$ are displayed for RR of a better outcome. The $\text{RR}^{+}$ has a very similar pattern to the RD. Both the PPO model and PO models have decreasing $\text{RR}^{+}$s as the value becomes more extreme. The $\text{ARR}^{+}$ (1.03, 95\% CI: 1.003 to 1.06) for the PPO model is the only `significant' result, while the $wRR^{+}$ (1.04, 95\% CI: 0.99 to 1.10) from the PPO model and the $\text{ARR}^{+}$ (1.02, 95\% CI: 0.996 to 1.05) and $wRR^{+}$ (1.04, 95\% CI: 0.99 to 1.09) from the PO model are all not significant. 
In panel C, the cumulative $\text{RR}^{-}$s for the individual outcomes, $\text{ARR}^{-}$, and $wRR^{-}$ are displayed for RR of a worse outcome. The $\text{RR}^{-}$ has a very similar pattern to the OR. The PPO model has increasing $\text{RR}^{-}$s as the value becomes more extreme. The $\text{ARR}^{-}$ (1.64, 95\% CI: 1.12 to 2.49) for the PPO model is the only significant result, while the $wRR^{-}$ (1.19, 95\% CI: 0.98 to 1.46) from the PPO model and the $\text{ARR}^{-}$ (1.21, 95\% CI: 0.97 to 1.51) and $wRR^{-}$ (1.17, 95\% CI: 0.97 to 1.42) from the PO model are all not significant. 

The $wRD$ and $wRR$ measures at day 28 are shown in Figure S3. 
The $wRD$ from the PPO is 0.04 (95\% CI: -0.01 to 0.08), and the $wRD$ from the PO model is 0.04 (95\% CI: -0.01 to 0.09). Both these estimates show a slight benefit of metformin, but they are not significant. 
The PPO model $wRR^{+}$ for a better outcome is 1.05 (95\% CI: 0.99 to 1.11), while the PO model $wRR^{+}$ is 1.06 (95\% CI: 0.99 to 1.12). These estimates are very similar to the day 14 estimates. 
The PPO model $wRR^{-}$ for a worse outcome is 1.21 (95\% CI: 1.00 to 1.47), while the PO model $wRR^{-}$ is 1.19 (95\% CI: 0.98 to 1.44). These estimates are very similar to the day 14 estimates. 


We carried out sensitivity analyses also adjusting for sex, BMI, and diabetes, in addition to the other study drugs and vaccination status, as these variables were slightly unbalanced across the metformin and placebo analysis groups; however, the results were very similar. We also carried out an analysis combining hypoxemia and nothing bad into one category because of concerns regarding the prevalence of spurious at-home hypoxemia measurements. These analyses show that metformin has a stronger effect on the more severe categories of this ordinal outcomes. Lastly, we conducted analyses using the optimal weights with the average of the treatment and control arm as derived by \cite{dg_clayton_odds_1974} instead of using just the probabilities from the control arm. We also used empirical plug in estimates for the weights based on the observed outcome distribution in the control arm. We did not find any major differences in these analyses.

\section{Discussion} \label{sec:discussion}
In this paper, we proposed a set of transparent population level summary measures for characterizing treatment effects on ordinal outcomes, along with efficient model-assisted Bayesian estimators based on a partial proportional odds model that facilitates covariate adjustment. 
Our approach provides a powerful method for summarizing ordinal outcomes in randomized controlled trials regardless of the validity of the proportional odds assumption, thereby effectuating the estimand framework. 
In particular, our approach shows negligible loss of power to the conventional proportional odds approach when the proportional odds assumption holds, and can exhibit substantial gains in power in non-proportional odds settings. 
Additionally the proposed summary measures are transparent with clearly defined weights for aggregating the treatment effects on each break point of the ordinal outcome.
Our proposed approach may be appealing for RCTs where one is unsure of the reasonableness of the PO assumption \textit{a priori} and since the analysis plan needs to be prespecified, it may not be reasonable to rely on such a strong assumption being met. 

The proposed summary measures reflect weighted arithmetic or geometric means with weights based on the outcome distribution in the control arm. These weights will reflect the trial population in a RCT of an experimental treatment against the standard of care and ensure the weights corresponding to each break point specific treatment effect do not depend on the magnitude of the treatment effect itself. 
In a sensitivity analysis, we confirmed that summary measures with equal weights can be highly variable when there are components with very low incidence. For example, the COVID-OUT trial protocol actually defined a binary composite with 7 components: hypoxemia, hypoxemia with supplemental oxygen, ED visit, hospitalization, hospitalization with ventilator support, hospitalization with ventilator support for three or more days, and death \citep{bramante_randomized_2022}. However, many of these components rarely occurred despite randomizing nearly 1,500 participants, e.g., no one was hospitalized with ventilator support for three or more days. Table S5 compares the summary measures for the adjusted analyses with the four level and seven level ordinal outcomes. The more granular outcome highlights that when there are rare outcomes, the proposed weighted average summary measures are more robust than the equally weighted measures. In particular, the $wRD$, $wOR$, $wRR^{+}$, and $wRR^{-}$ are similar for the four level and seven level ordinal outcomes at days 14 and 28, while the ARD, AOR $\text{ARR}^{+}$ and $\text{ARR}^{-}$ have substantially more variability with seven levels compared with four levels. 

We only considered model-assisted estimators based on a partial proportional odds model, which are highly flexible. Another option would be to use a constrained partial proportional odds model that allows imposing constraints on how the log odds ratio can change (or not change) across the break points of the outcome \citep{peterson_partial_1990}. Such constraints may provide further efficiency gains when they are in fact correct. In some settings, this may offer an effective middle ground between the proposed approach and a fully proportional odds model. 

Our proposed population level summary measures and their model-assisted Bayesian estimators characterize treatment effects on ordinal outcomes with transparency without making rigid parametric assumptions like proportional odds. Our approach tends to be as or more powerful than the conventional proportional odds model approach while also offering greater transparency regarding the aggregation of treatment effects across the various break points of the ordinal outcome. 
Our weights are solely based on incidence of the outcome, but one could consider using other weighting schemes that emphasize the outcome components with greater clinical relevance, like hospitalization and death in the COVID-OUT trial. 
Although we proposed weighted mean risk difference and weighted geometric mean risk ratio measures, we generally favor the proposed weighted geometric mean odds ratio measure. 
The risk ratio measures lack invariance to whether the outcome is ordered from best to worst or vice versa, and they can be sensitive to this arbitrary choice when the best and worst outcome categories have low incidence.
The weighted risk difference measure may appeal from an interpretation perspective, but is limited by the difficulty of aggregating cumulative risk differences with potentially widely varying incidence.
In settings where the best and worst outcome categories have high incidence such that the hypothesized risk difference may be relatively constant across the outcome break points, then a weighted mean risk difference may be an appealing population level summary measure for interpretability.
However, this was not the case in the motivating COVID-OUT trial where hospitalization and death were rare. 

The COVID-OUT trial data analyzed in Section 4 is available via from the Data Repository for the University of Minnesota (DRUM) \url{https://hdl.handle.net/11299/269790}. R programs to reproduce the simulation studies reported in Section 3 are available from the first author's GitHub page \url{https://github.com/lindseyturner/ordinal_summary_measures}.

\section*{Acknowledgements}
Drs. Murray and Bramante received support from the Parsemus Foundation and Rainwater Charitable Foundation. Dr. Bramante was also supported by grants (KL2TR002492 and UL1TR002494) from the National Center for Advancing Translational Sciences (NCATS) of the National Institutes of Health (NIH) and by a grant (K23 DK124654) from the National Institute of Diabetes and Digestive and Kidney Diseases of the NIH.

\bibliographystyle{biom} \bibliography{references}

\renewcommand{\thetable}{S\arabic{table}}
\renewcommand{\thefigure}{S\arabic{figure}}

\renewcommand{\thesection}{\Alph{section}}




\clearpage

\begin{center}
    {\bfseries\Large{Supplementary Material}}
\end{center}

\setcounter{section}{0}
\renewcommand{\thesection}{\Alph{section}}

\section{Additional Simulation Results}

\begin{figure}[H]
 \centerline{\includegraphics[width=6in]{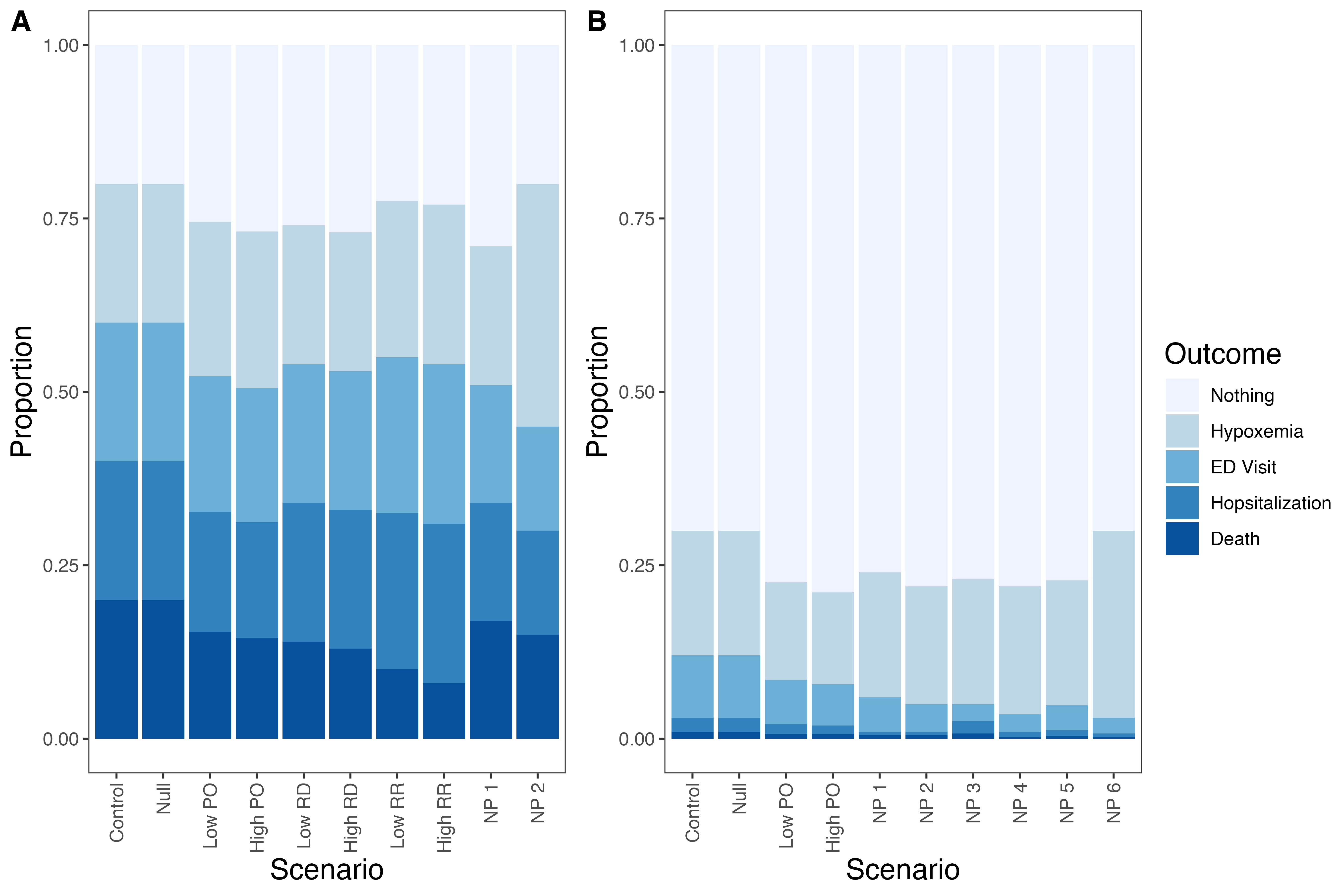}}
\caption{Distribution of the ordinal outcome in each of the simulation settings. Panel A shows setting 1 where the ordinal outcome is evenly distributed in the control arm. Panel B shows setting 2 where the ordinal outcome is distributed mimicking the COVID-OUT distribution.}
 \label{f:sim_dist}
\end{figure}

Below we report additional results from the simulation studies. 
The distribution of the control and treatment arm in each of the simulation scenarios is depicted in Figure \ref{f:sim_dist}. Table \ref{t:evenpower_true} displays the true values for the $wOR$ summary measures with the different possible weighting schemes. The true value for a PO model is estimated with a large sample size of 1,000,000. The $wOR$ with overall weights most closely estimates the same value as a proportional odds model. However, the power when using the overall weights is similar to the power when using the control weights.

\begin{table}[H]
\centering
 \def\~{\hphantom{0}}

\resizebox{\columnwidth}{!}{%
\begin{tabular}{>
{\centering\arraybackslash}p{4em}ccccccc}
\toprule
& & \multicolumn{3}{c}{Control Weights} 
& \multicolumn{3}{c}{Overall Weights} \\
\cmidrule(l){3-5}\cmidrule(l){6-8}
Scenario & PO & wOR & wOR cumulative & wOR sum & wOR & wOR cumulative & wOR sum\\
\midrule
Null & 0 & 0 & 0 & 0 & 0 & 0 & 0\\
Low PO & 0.31 & 0.31 & 0.31 & 0.31 & 0.31 & 0.31 & 0.31\\
High PO & 0.38 & 0.39 & 0.39 & 0.39 & 0.39 & 0.39 & 0.39\\
Low RD & 0.29 & 0.3 & 0.3 & 0.32 & 0.3 & 0.3 & 0.32\\
High RD & 0.36 & 0.36 & 0.36 & 0.37 & 0.35 & 0.35 & 0.37\\
Low RR & 0.32 & 0.35 & 0.35 & 0.37 & 0.32 & 0.33 & 0.36\\
High RR & 0.38 & 0.44 & 0.44 & 0.47 & 0.38 & 0.4 & 0.44\\
NP 1 & 0.34 & 0.33 & 0.33 & 0.33 & 0.34 & 0.33 & 0.34\\
NP 2 & 0.39 & 0.38 & 0.38 & 0.35 & 0.38 & 0.39 & 0.34\\
\bottomrule
\end{tabular}}
\caption{True Values for Setting 1 - Evenly Distributed Outcome, showing values for both the control weights as well as the overall weights (defined as the average of the control and treatment weights).}
\label{t:evenpower_true}
\end{table}

Table \ref{t:evenpower_sup} shows the type I error rate and power the other summary measures ($wOR$ Cumulative, $wOR$ Sum, ARD, $wRD$, $ARR^{+}$, $wRR^{+}$, $ARR^{-}$, and $wRR^{-}$). The type I error rate is approximately constant at 5\% across all levels of the outcome. The $wOR$ measures tend to have similar or higher power than the PO model, besides in the NP 2 scenario. When PO holds, all weighted summary measures have similar power. In certain scenarios, the $wRR^{+}$ has the highest power, while in other scenarios the $wRR^{-}$ has the highest power.

\begin{table}
\centering
 \def\~{\hphantom{0}}

\resizebox{\columnwidth}{!}{%
\begin{tabular}{>
{\centering\arraybackslash}p{4em}ccccccccc}
\toprule
& \multicolumn{9}{c}{Ordinal Composite} \\
\cmidrule(l){2-10}
Scenario & PO & wOR cumulative & wOR sum & ARD & wRD & $ARR^{+}$ & $wRR^{+}$ & $ARR^{-}$ & $wRR^{-}$\\
\midrule
Null & 0.06 & 0.06 & 0.06 & 0.06 & 0.06 & 0.05 & 0.05 & 0.05 & 0.05\\
Low PO & 0.80 & 0.80 & 0.79 & 0.80 & 0.79 & 0.78 & 0.78 & 0.75 & 0.78\\
High PO & 0.92 & 0.93 & 0.92 & 0.93 & 0.92 & 0.91 & 0.91 & 0.89 & 0.91\\
Low RD & 0.75 & 0.76 & 0.79 & 0.76 & 0.71 & 0.75 & 0.72 & 0.79 & 0.76\\
High RD & 0.88 & 0.89 & 0.91 & 0.89 & 0.86 & 0.87 & 0.84 & 0.90 & 0.89\\
Low RR & 0.79 & 0.85 & 0.88 & 0.82 & 0.74 & 0.59 & 0.56 & 0.98 & 0.94\\
High RR & 0.93 & 0.97 & 0.97 & 0.95 & 0.88 & 0.75 & 0.71 & 1.00 & 0.99\\
NP 1 & 0.86 & 0.84 & 0.85 & 0.84 & 0.86 & 0.91 & 0.91 & 0.69 & 0.74\\
NP 2 & 0.92 & 0.92 & 0.85 & 0.93 & 0.95 & 0.69 & 0.76 & 0.91 & 0.95\\
\bottomrule
\end{tabular}}
  \caption{Power for Setting 1 - Evenly Distributed Outcome. All columns show the power for an ordinal outcome using a PO model, wOR with the cumulative weights, wOR with the sum weights, ARD, wRD, $\text{ARR}^{+}$, $wRR^{+}$, $\text{ARR}^{-}$, and $wRR^{-}$.}
\label{t:evenpower_sup}
\end{table}

In Table \ref{t:covidpower_true}, the true values for setting 2 are displayed. The $wOR$ with overall weights most closely estimates the PO model. Both the $wOR$ cumulative and $wOR$ sum are estimating a larger value than the PO model in all scenarios when the PO assumption does not hold. 
Table \ref{t:covidpower_sup} displays the power for setting 2 mimicking COVID-OUT. Again, there are very similar results with type I error rate being constant around 5\%. The unweighted average summary measures tend to have much more variability in the power. 
The $wRD$ and $wRR^{+}$ tend to have very similar power, while the $wRR^{-}$ tends to have a slightly higher power. 
The $wOR$ cumulative and $wOR$ sum tend to have a higher power driven by estimating a different value than the OR from PO model.

\begin{table}[H]
\centering
 \def\~{\hphantom{0}}

\resizebox{\columnwidth}{!}{%
\begin{tabular}{>
{\centering\arraybackslash}p{4em}ccccccc}
\toprule
& & \multicolumn{3}{c}{Control Weights} 
& \multicolumn{3}{c}{Overall Weights} \\
\cmidrule(l){3-5}\cmidrule(l){6-8}
Scenario & PO & wOR & wOR cumulative & wOR sum & wOR & wOR cumulative & wOR sum\\
\midrule
Null & 0 & 0 & 0 & 0 & 0 & 0 & 0\\
Low PO & 0.38 & 0.39 & 0.39 & 0.39 & 0.39 & 0.39 & 0.39\\
High PO & 0.46 & 0.47 & 0.47 & 0.47 & 0.47 & 0.47 & 0.47\\
NP 1 & 0.35 & 0.38 & 0.52 & 0.48 & 0.36 & 0.49 & 0.45\\
NP 2 & 0.47 & 0.5 & 0.64 & 0.6 & 0.47 & 0.61 & 0.57\\
NP 3 & 0.41 & 0.44 & 0.52 & 0.47 & 0.41 & 0.49 & 0.46\\
NP 4 & 0.5 & 0.55 & 0.77 & 0.69 & 0.5 & 0.71 & 0.65\\
NP 5 & 0.43 & 0.46 & 0.62 & 0.56 & 0.44 & 0.58 & 0.54\\
NP 6 & 0.13 & 0.22 & 0.6 & 0.46 & 0.14 & 0.45 & 0.42\\
\bottomrule
\end{tabular}}
  \caption{True Values for Setting 2 - Mimicking COVID-OUT, showing values for both the control weights as well as the overall weights (defined as the average of the control and treatment weights).}
\label{t:covidpower_true}
\end{table}

\begin{table}[H]
\centering
 \def\~{\hphantom{0}}

\resizebox{\columnwidth}{!}{%
\begin{tabular}{>
{\centering\arraybackslash}p{4em}ccccccccc}
\toprule
& \multicolumn{9}{c}{Ordinal Composite} \\
\cmidrule(l){2-10}
Scenario & PO & wOR cumulative & wOR sum & ARD & wRD & $\text{ARR}^{+}$ & $wRR^{+}$ & $\text{ARR}^{-}$ & $wRR^{-}$\\
\midrule
Null & 0.04 & 0.05 & 0.05 & 0.04 & 0.04 & 0.04 & 0.04 & 0.06 & 0.04\\
Low PO & 0.78 & 0.77 & 0.75 & 0.71 & 0.78 & 0.74 & 0.78 & 0.27 & 0.79\\
High PO & 0.91 & 0.90 & 0.89 & 0.87 & 0.91 & 0.88 & 0.91 & 0.36 & 0.92\\
NP 1 & 0.71 & 0.92 & 0.87 & 0.86 & 0.64 & 0.84 & 0.63 & 0.58 & 0.78\\
NP 2 & 0.91 & 0.98 & 0.97 & 0.94 & 0.87 & 0.94 & 0.87 & 0.66 & 0.94\\
NP 3 & 0.82 & 0.93 & 0.87 & 0.89 & 0.78 & 0.88 & 0.77 & 0.34 & 0.87\\
NP 4 & 0.93 & 1.00 & 0.99 & 0.99 & 0.89 & 0.99 & 0.88 & 0.86 & 0.97\\
NP 5 & 0.86 & 0.97 & 0.96 & 0.96 & 0.81 & 0.95 & 0.79 & 0.70 & 0.91\\
NP 6 & 0.15 & 0.94 & 0.80 & 0.77 & 0.08 & 0.59 & 0.07 & 0.85 & 0.44\\
\bottomrule
\end{tabular}}
  \caption{Power for Setting 2 - Mimicking COVID-OUT. AAll columns show the power for an ordinal outcome using a PO model, wOR with the cumulative weights, wOR with the sum weights, ARD, wRD, $\text{ARR}^{+}$, $wRR^{+}$, $\text{ARR}^{-}$, and $wRR^{-}$.}
\label{t:covidpower_sup}
\end{table}

\newpage

\section{Additional COVID-OUT Analyses}

\begin{figure}[H]
 \centerline{\includegraphics[width=6in]{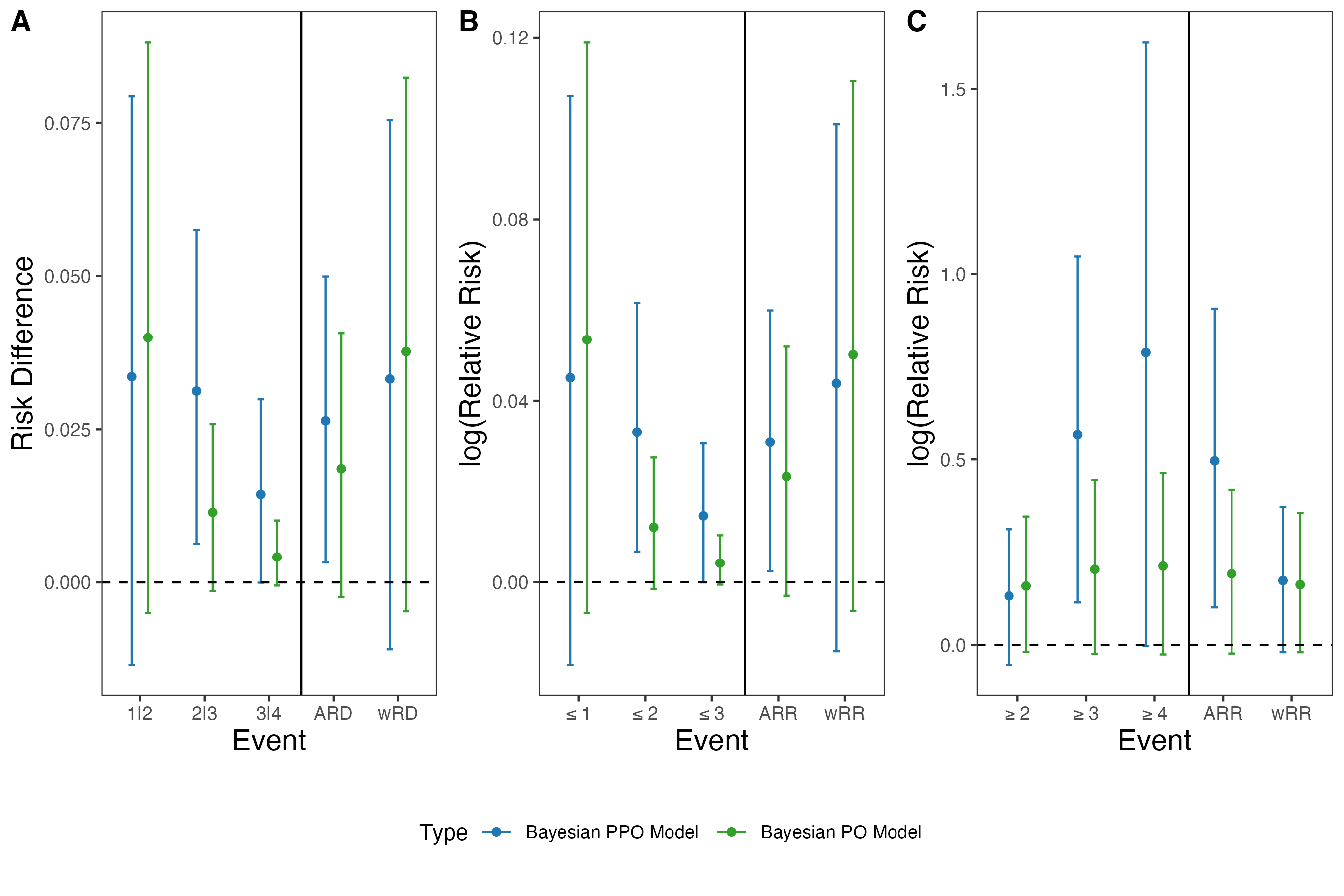}}
\caption{Adjusted analyses for COVID-OUT dataset at day 14 for metformin relative to placebo. Panels A, B, and C show the RD, log($\text{RR}^{+}$), and log($\text{RR}^{-}$). First showing the individual measures when dichotomizing the outcome and then the unweighted average and weighted average measures for the ordinal outcome for both a Bayesian partial proportional odds (PPO) model (blue) and a Bayesian proportional odds (PO) model (green) using the Bayesian bootstrap for marginalization. The events are defined as nothing bad (1), hypoxemia (2), ED visit (3), and hospitalization/death (4).}
 \label{f:adj14_supp}
\end{figure}

\begin{figure}[H]
 \centerline{\includegraphics[width=6in]{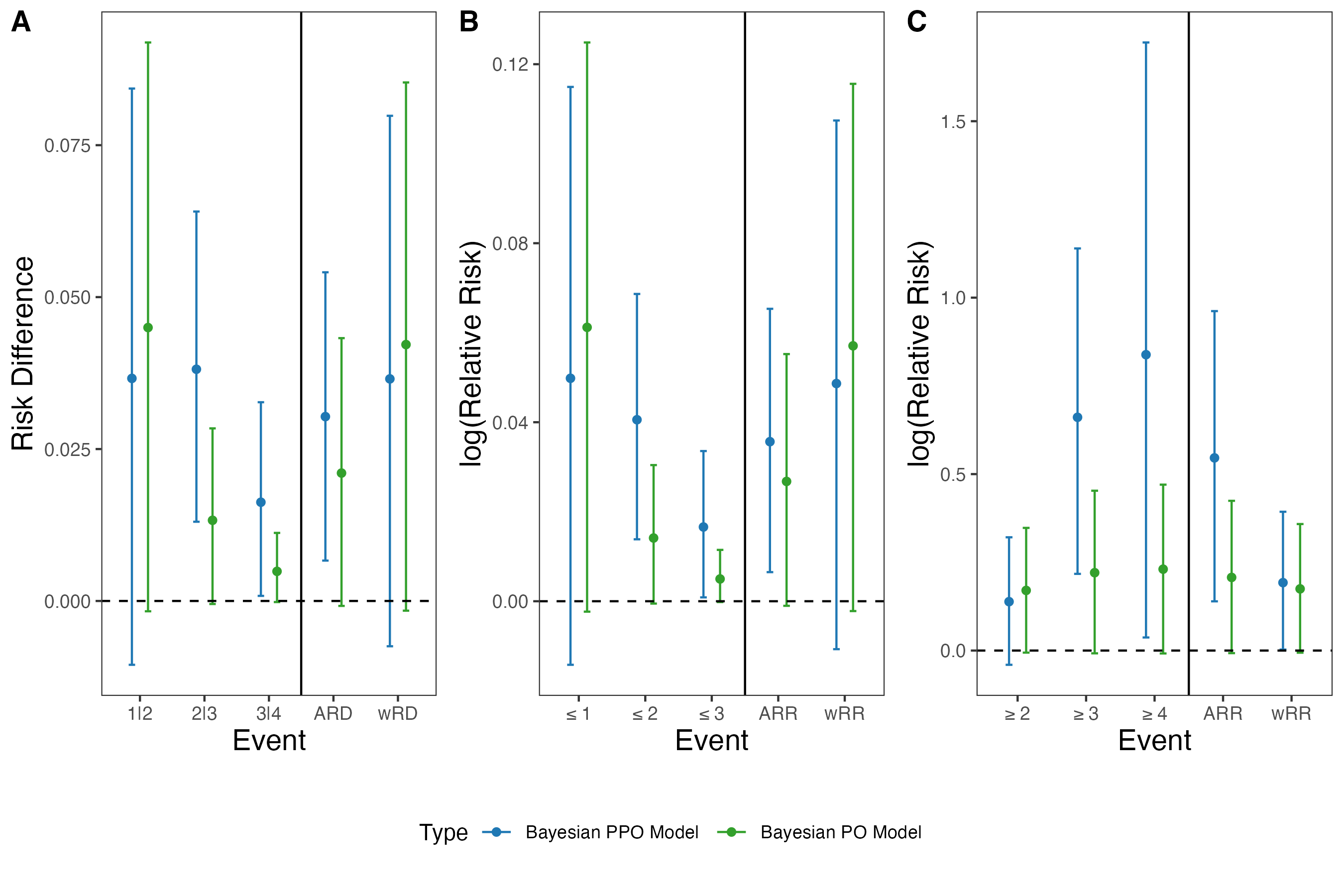}}
\caption{Adjusted analyses for COVID-OUT dataset at day 28 for metformin relative to placebo. Panels A, B, and C show the RD, log($\text{RR}^{+}$), and log($\text{RR}^{-}$). First showing the individual measures when dichotomizing the outcome and then the unweighted average and weighted average measures for the ordinal outcome for both a Bayesian partial proportional odds (PPO) model (blue) and a Bayesian proportional odds (PO) model (green) using the Bayesian bootstrap for marginalization. The events are defined as nothing bad (1), hypoxemia (2), ED visit (3), and hospitalization/death (4).}
 \label{f:adj28_supp}
\end{figure}

Below we report results for the other proposed population level summary measures. 
The results for day 14 are presented in Figure \ref{f:adj14_supp}. In panel A, the cumulative RD is shown for each level of the outcome as well as the ARD and wRD. 
In panel B, the cumulative $\text{RR}^{+}$ is shown for each level of the outcome as well as the $\text{ARR}^{+}$ and $wRR^{+}$. 
In panel C, the cumulative $\text{RR}^{-}$ is shown for each level of the outcome as well as the $\text{ARR}^{-}$ and $wRR^{-}$. 
These plots show the estimates from both a Bayesian partial proportional odds model as well as from a Bayesian proportional odds model. 
Similarly, Figure \ref{f:adj28_supp} shows the same summary measures at day 28. These results are similar to the results from day 14.

As a sensitivity analysis, we analyzed the ordinal outcome using the original seven levels.
Due to the rareness of the outcome, when fitting the model with the rmsb package in R, we set priorsdppo equal to 2.5 for all models (both the four level and seven level outcomes) to ensure every event had a non negative probability. If an event had a negative probability, the probability was truncated at zero and the probabilities were rescaled to sum to one. The priorsdppo parameter penalizes the partial proportional odds term. All models were adjusted for the other study drugs (fluvoxamine and ivermectin) as well as COVID-19 vaccination status. The more granular outcome highlights that when there are rare outcomes, the weighted average summary measures are more robust. The weighted average summary measures are very similar between the four level and seven level outcomes, while the average summary measure has more variability between the four and seven level outcomes.

\begin{table}[H]
\centering
\resizebox{\columnwidth}{!}{%
\begin{tabular}[t]{p{1in}p{1in}p{1in}p{1in}p{1in}p{1in}p{1in}p{1in}p{1in}}
\toprule
  & AOR & wOR & ARD & wRD & $\text{ARR}^{+}$ & $wRR^{+}$ & $\text{ARR}^{-}$ & $wRR^{-}$\\
\midrule
Day 14: 4 level & 0.22 (-0.06, 0.49) & 0.19 (-0.06, 0.43) & 0.02 (0, 0.04) & 0.03 (-0.01, 0.08) & 0.02 (-0.01, 0.05) & 0.04 (-0.02, 0.1) & 0.2 (-0.05, 0.45) & 0.14 (-0.05, 0.32)\\
Day 14: 7 level & 0.22 (-0.06, 0.49) & 0.18 (-0.07, 0.43) & 0.01 (0, 0.03) & 0.03 (-0.01, 0.08) & 0.01 (0, 0.03) & 0.05 (-0.02, 0.11) & 0.2 (-0.05, 0.45) & 0.14 (-0.05, 0.33)\\
\hline
Day 28: 4 level & 0.27 (-0.01, 0.54) & 0.23 (-0.04, 0.48) & 0.02 (0, 0.05) & 0.04 (-0.01, 0.09) & 0.03 (0, 0.06) & 0.06 (-0.01, 0.12) & 0.24 (-0.01, 0.49) & 0.17 (-0.03, 0.36)\\
Day 28: 7 level & 0.25 (-0.01, 0.51) & 0.22 (-0.02, 0.47) & 0.01 (0, 0.03) & 0.04 (0, 0.09) & 0.02 (0, 0.04) & 0.06 (-0.01, 0.12) & 0.23 (-0.01, 0.48) & 0.17 (-0.01, 0.36)\\
\bottomrule
\end{tabular}
}
\centering
\caption{Adjusted COVID-OUT Sensitivity Analysis. This table shows the point estimate and 95\% credible interval for the ARD, wRD, AOR, wOR, $ARR^{+}$, $wRR^{+}$, $ARR^{-}$, and $wRR^{-}$ for the four level and seven level ordinal outcomes at days 14 and 28.}
\label{tab:suptab}
\centering
\end{table}

\label{lastpage}

\end{document}